# CELIO: An Application Development Framework for Interactive Spaces

**Yedendra B. Shrinivasan**
IBM Research, USA
yshrini@us.ibm.com

**Yunfeng Zhang**
IBM Research, USA
zhangyun@us.ibm.com

## ABSTRACT
Developing applications for interactive space is different from developing cross-platform applications for personal computing. Input, output, and architectural variations in each interactive space introduce big overhead in terms of cost and time for developing, deploying and maintaining applications for interactive spaces. Often, these applications become on-off experience tied to the deployed spaces. To alleviate this problem and enable rapid responsive space design applications similar to responsive web design, we present CELIO application development framework for interactive spaces. The framework is micro services based and neatly decouples application and design specifications from hardware and architecture specifications of an interactive space. In this paper, we describe this framework and its implementation details. Also, we briefly discuss the use cases developed using this framework.

### Author Keywords
Large displays, Interactive Spaces, Public displays, Intelligent room

### ACM Classification Keywords
H.5.m. Information interfaces and presentation (e.g., HCI): Miscellaneous.

## INTRODUCTION
Interactive spaces are conceived in various forms from simple interactive displays in public spaces [9], ubiquitous computing spaces [6,10], to smart rooms [1,2,11] that are supported by intelligent agents, surrounded by displays and enabled with different interaction capabilities. An interactive space is created as a network of devices and objects compared to the single device paradigm in personal computing devices such as laptop or smartphones. In single device paradigm, the system enables a single user at a time, with other devices added on-demand to the primary device through universal plug and play interfaces. Interactive spaces are used by a group of users working in single context or multiple groups with different contexts at the same time, for instance, break-out sessions. The context is abstractly defined by users' tasks, intelligent agents, micro-services, and data used by a group. The design elements in each interactive space vary in terms of its physical constituents, such as doors, lights and tables, input capabilities, sensors, actuators and displays. Hence, an application designed to work in an interactive space cannot be used in another space due to these variations in architecture and hardware.

In this paper, we present the CELIO (Cognitive Environments Library for Input/Output) architecture and API specification to support I/O-resource-independent application development for interactive spaces varying in architecture and hardware. First, we present the motivating scenarios as background and infer the system design requirements based on the lessons learned. Next, we present the CELIO architecture and interweave that with implementation details based on the state-of-the-art distributed messaging and web technologies. Finally, we discuss the API framework and illustrate it through the development of universal remote and proxemics applications for the interactive space in our lab.

## MOTIVATION
Prior work by [16] described three interactive spaces for commercial environments with different space architectures and hardware setup. The first interactive space is an intelligent room that supports multi-modal interaction using speech, gesture, 3D pointing, and static and moving displays. This space is primarily used for data exploration and had around three applications (for different customers). The speech is captured using either table microphone, users' smartphones, or a centralized microphone array and commands are parsed and dispatched to corresponding application agents. Kinect™ is used to identify user movements, pointing directions, and gestures. Most of the visualizations on display are created using web technologies and accepted mouse, touch and keyboard events. Speech and Kinect events are indirectly consumed by the applications through an application executor in the backend application server. The space also has capability to produce spatialized audio output (with speakers placed at different corners of the room). The second interactive space is a customer experience center that had large display wall (40ft), interactive surfaces and tablets along with 3D pointing using Oblong[1]'s Mezzanine™ wand system. The main user of this space is a docent who presents interactive digital content and guides visitors during self-exploration sessions. The third interactive space is a cluster of five interactive spaces similar to the second setting, but with a smaller display wall (~15ft). The critical reason why the authors were unable to easily reuse the applications and each space stayed with one-off experience is not only due to the fact of space and hardware variations, but also the lack of developer enablement to manage these variations. This resulted in a lot of reengineering of the software deployed in each space, which became an overhead to maintain. Similar issues came up during web evolution where developing webpages for different screen resolutions and different devices became an overhead, but was somewhat resolved through different responsive web UI design frameworks. Even though, each interactive space needs rethinking on content, visual and interaction design as discussed in [4,8,19], the management of middleware and application development for interactive spaces should not be an overhead. The development process should be as consistent and close as responsive web development experience.

---
[1] http://www.oblong.com

## SYSTEM DESIGN REQUIREMENTS

An application framework for interactive spaces has inherently different design requirements than that for personal computing environments. In personal computing, a single device such as a PC, a laptop, or a smart phone has all the computational power and hardware it needs to communicate with its user. A personal computing device has little knowledge about other surrounding personal computing devices unless they are connected via a universal plug and play interface, and applications are mostly designed to run on a single device for a single user. In interactive space, there are many different sensors, computing platforms, displays, robots, and other intelligent objects. Applications designed for such a space needs to span across multiple platforms and accommodate multiple users. To meet this goal, an application framework for intelligent spaces needs to satisfy several requirements discussed next.

### Support development of distributed systems

An interactive space often has heterogeneous platforms driving different hardware. For example, in the first interactive space -- the cognitive environment lab discussed in the previous section, a Linux system is used to drive a display wall due to the flexibility in customizing the Linux graphics layer, but a Mac was used to drive our audio system because of its better support for multichannel audio hardware. In this case, an application written for such an environment needs to somehow reside on both platforms in order to use both video and audio hardware. In addition, the sensors and actuators used in the space often come with their own platforms, and to consume sensor data and send commands to the actuators, the applications again need to be written in a distributed manner.

Though developing distributed systems introduces many challenges, including openness, security, failure handling, concurrency, transparency, and service discovery [3], but it also brings many benefits. A distributed system is inherently a parallel system, and thus system developers can efficiently take advantage of all the computational power present in a room and in the cloud. A distributed system often has to be loosely coupled, and therefore its components has to be very modular. As a result, developers can more effectively develop different components at the same time, and that the architecture as a whole is more tolerant to changes.

### Support spatial awareness of things, users, and devices

An interactive space is conceived with a spatial reference frame to support location of things, displays, and users. The spatial reference frame supports various interaction techniques such as proxemics, 3D pointing, user tracking, device tracking, gesture interpretation, pan-tilt-zoom cameras, and robot navigation. In personal computers, only displays have a coordinate system. Pointing data such as movements of a mouse and leap motion gestures are always projected to the display coordinate system to interact with objects on the display. Few spaces support proxemics interaction by adapting to the user's distance from the display. In an interactive space, measuring the absolute location of things and users can significantly enhance user experience: audio could be played from a speaker next to the user, and the user can point to things to activate them or to change their behavior. Also, users can define their hotspots in space and bookmark a digital content in space. For instance, tapping/pointing a notice board can result in some actions such as playing music or displaying a presentation. Pointer devices such as HTC[2] Vive™, Oblong wand, and Kinect will have their own spatial reference system. It is important to transform pointing data from different spatial reference system to a selected spatial reference system for consistent spatial interaction.

### Support multi-user interactions

Unlike personal computing, an interactive space is meant to be used by a group of users at the same time. This fundamental difference causes the traditional graphic user interfaces unsuitable for intelligent spaces. Such incompatibility can be particularly seen in how PCs handle input. For example, although multiple keyboards and mice can be plugged in to a PC, the keyboards would share the same input point and the mice would share the same cursor, so that two users cannot type on the same document on a PC simultaneously. Similarly, although some large displays support multi-touch, they do not differentiate touch points from different users, which makes it difficult for users to efficiently collaborate using a large surface. In an interactive space, users need to simultaneous interact on different parts of the display either using pointing device, gestures or speech. Therefore, a new approach to handling input devices is needed for applications to support multiple users.

---

[2] https://www.htcvive.com/us/

**Support inter-space communication**
To support real-time remote collaboration across different interactive spaces, in addition to sharing view and audio, users can interact with visual objects using pointing and speech. An interactive space can stream 360º view of the room or contents from the display to remote locations. In UbiHostpot [14, users can simultaneously interact with hosted contents and share their contents in public displays across different locations in a city. In [17], remote users using virtual reality devices such as Oculus Rift™ can collaboratively interact with displays and things in the room. They can select and move the contents on display using virtual pointing, or use speech to drive application in the space from remote location in combination with virtual pointing. While supporting such inter-space communication, remote users' presence needs to be explicitly referenced and made aware of to consistent user interaction and content negotiation.

**Support the Common User Interface Requirements**
Beside the above requirements that are unique to interactive spaces, the application framework for IS also needs to satisfy some conventional requirements in personal computing such as enabling real-time interaction and device plug and play. Users have different expectations about the application response latency for different interaction paradigms, and for some interactions such as pointing, users typically expect the latency to be very short. The application framework thus should not cause long delays. This may be a problem in a distributed computing environment and thus should be carefully addressed in designing the framework. Plug and play is another important aspect that an application framework should support. In an interactive space, mobile devices may come and leave frequently, and the framework should enable the users to use such devices without arduous setup processes each time a device is brought into the environment. In particular, the framework should support users using their own mobile devices as input/output endpoints for the intelligent space. This would immediately add many input/output devices to the intelligent environment and allow novice users who are not yet familiar with the devices in the environment to comfortably interact with it.

## CELIO ARCHITECTURE
We propose an application framework for interactive spaces, which we call the cognitive environment library for input/output (CELIO). Two main goals drive the design of CELIO: (1) To support the interactive spaces system design requirement as described in the previous section; (2) to enable application developers to easily develop and deploy applications that make full use of the spaces' intelligent sensing, ubiquitous displays, and other actuators. To meet these goals, we adopted a micro service-oriented distributed architecture. Since some of our architectural decisions are based on the current state-of-the-art technologies for message broker and displays based on desktop-web hybrid development environment, we interweave architectural discuss with the implementation details. Figure 1 shows the components of the CELIO architecture. The architecture consists of hardware workers, a message broker, and a registry. Generally speaking, the workers send events to the applications through the broker, and the applications send commands to the workers through messaging and REST calls. This section describes for each component the architectural decisions we made, the API, and the implementation details.

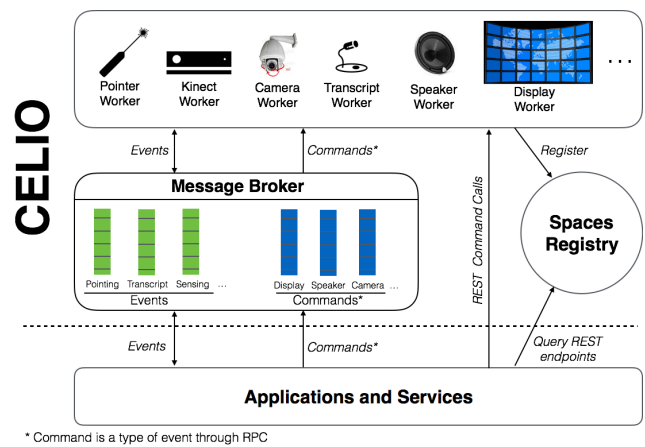

Figure 1. The components of CELIO architecture.

**Spaces Registry**
A spaces registry is a federation of registries of connected interactive spaces. Each interactive space has a local registry to which all hard-wired services (workers) are registered. Micro-services such as intelligent analytics agents, user profiles, and data services are mostly registered in the spaces registry, unless, security concerns require these services to be in only specific interactive spaces registries. In addition to record keeping, it has two micro-services to enable discovery of other micro-services by type, name or location and to manage lifecycle of all micro-services.

**Message Broker**
At the heart of the CELIO architecture is the message broker. A worker publishes a certain type of events to the broker, which then routes the events to the applications that subscribe to that type of events. This publish-subscribe pattern is extremely suitable for writing event processing programs, and since user interface programming is mostly about processing input events, this pattern is also well suited for writing interactive applications.

In the past research, message brokers have been used to manage content on multiple displays in public and private spaces along with support for interaction via smartphones. MagicBroker [5] uses a centralized message broker to enable user interaction and content sharing on public displays using smartphones. A smartphone connects to the system via QRCode, SMS or VoiceXML™. In UbiBroker [7], a message broker is used to synchronize application and public displays' states across multiple users in multiple locations in a city. Naber et al. [13] uses a message broker to provide an application interface for content creation for large public displays as well as content sharing between public and private displays. None of these architectures supports multi-modal interaction or enable manipulation of things in space and contents on displays by different input mechanisms and multiple users. Only Nabel et al. attempted to provide an application provider interface to enable developers to write content management applications for interactive public spaces.

The reason why we use a centralized message broker as opposed to having the applications subscribe to the workers directly is threefold: (a) It enables device plug and play. Without a message broker, an application needs to somehow know that a new device is connected in order to connect to it and subscribe to its events. This device discovery usually requires additional middleware. However, with a message broker, such middleware is no longer needed, because the applications will simply start receiving events from the device worker whenever it is connected to the broker. This dramatically simplifies the architecture. (b) A message broker allows a variety of subscription methods, which satisfies many different application needs. Many message broker implementations such as Kafka™[3] and RabbitMQ™[4] allow wildcard subscriptions, e.g. *.pointing, which will subscribe to all events whose type ends with pointing. This enables applications to subscribe to a class of events from different sources and also allows applications to just subscribe to a more specific type if they need to. (c) A message broker eases security management because security measures such as authentication and message encryption only need to be implemented for the single broker, whereas if the applications subscribe to the workers directly, security measures need to be taken for each publishing endpoint.

In addition to sending events, we also use the message broker to send a portion of the commands that do not require responses. This way, the applications can issue commands without knowing where the workers are hosted. For commands that require responses (such as getting a window ID after creating a window on the display), we currently use REST (representational state transfer) calls because REST defines clear semantics for remote procedural calls and returning responses. Response-required remote procedural calls are typically hard to implement on a message broker because the broker does not permit the caller to talk to the callee directly. As a result, the caller would not know whether the call is actually handled by any services. The downside of using REST calls is that the caller needs to know the callee's address, and the callee's address may change often in a dynamic network environment. To address this issue, we created the space registry to maintain the REST services' addresses, and the caller needs to query the callee's address before making a call.

Adopting a message broker has some drawbacks, but they are largely alleviated by today's message broker systems. The drawbacks include the additional delay in sending a message because of the extra hop at the message broker, a potential single point of failure, and the potential overloading of the message broker. The first issue is very minor in the context of intelligent rooms because the message broker can be deployed locally, close to the workers. In our test with a RabbitMQ message broker, 99.5% of the messages were received with less than 2 ms delay. Such delay is acceptable for real-time human computer interactions. The second and the third issues can be addressed by running a cluster of message brokers to provide redundancies and load balancing, and several message broker implements supports clustering out of the box.

*Message Broker In CELIO*
In the CELIO architecture, we use the RabbitMQ message broker. It is developed based on the advanced message queuing protocol (AMQP) but it also supports several other protocols such as MQTT and STOMP. This wide range support allows RabbitMQ to span across different platforms that do not support AMQP, such as low powered sensor networks (via MQTT) and web clients (via STOMP). Because these message protocols are language neutral, libraries for RabbitMQ are available in many programming languages.

RabbitMQ supports many different message routing paradigms including direct (routing by topic names without wildcard matching), topic (routing by topic names with wildcard matching), and fan-out (no topic routing; every message is broadcasted to every subscriber). We primarily use its topic routing paradigm, because of its support for wildcard matching in topic subscription. As mentioned earlier, this feature eliminates the need for middle-layer managers described in some previous architectural proposals (such as [6]).

RabbitMQ also supports many different message distribution paradigms, which are useful in supporting many different scenarios. For example, by default, we send the messages to each subscriber asynchronously, but in some instances, we allow the subscriber to fetch a message from the message queue whenever needed. This is useful in the text-to-speech scenario, where multiple sentences may be queued and we want the speaker worker to say the sentences one by one. Using the default subscription paradigm, the speaker would receive a sentence before finish saying the previous one, causing it to cut off the previous utterance. With the fetch paradigm, the speaker worker can process the messages at its own pace. Another feature of RabbitMQ is that if multiple applications await in the same queue, they can process messages in a round-robin fashion. This is useful for distributing heavy workload such as face recognition among several services. Finally, RabbitMQ also supports extensions to allow developer to create their own message distribution paradigm. One such useful extension is a history cache exchange which caches recent messages and send them to a client when the client first connects to the broker. This is useful for example in notifying the clients about the state of the network such as which workers are present when a client first connects.

---
[3] http://kafka.apache.org

[4] https://www.rabbitmq.com

**Input Workers**
The workers can be thought of as drivers of CELIO. They usually run on the machines that have hardware connections to the sensors or actuators. They publish events from the hardware and accept commands sent from the applications. To support multimodal input, we have created multiple workers that support speech and gestural interactions. This section elaborates on these input workers, provides an overview of the application programming interfaces they support, and discusses the implementation details.

*Transcript Workers*
The main responsibility of a transcript worker is to capture audio from the microphones, send the audio stream to a speech-to-text (STT) transcription service, and publishes the transcription to the message broker. The transcription are then further processed by multiple natural language processing services to perform named entity recognition, intent recognition, and dialog interactions. There are many commercial STT services available today, and we use the Watson STT[5] service in our implementation. Watson STT provides some useful features such as keyword spotting and smart formatting in addition to accurate transcription (6.9% word error rate on the switchboard benchmark [15]). Keyword spotting allows the developer to specify keywords that should be reported even if they are not in the top transcription results. This is useful, e.g., for attention word capturing. Smart formatting converts numbers, dates, and times into certain standard formats so that they can be easily parsed by down-stream applications.

Our transcript worker supports multi-channel transcription by spawning a STT session for each audio channel. This is useful when multiple microphones are connected to the same computer. Without multi-channel transcription, different users' voices will be mixed together and the transcription will be a mix of many speech streams as well. With multi-channel transcription, a channel can also be identified by speaker identification programs or by simply saying "This is Jon speaking". This helps multimodal fusion when we try to fuse speech and gesture commands for multiple users.

Our transcript worker also pauses listening on far-range microphones when the avatar itself is talking. In the CELIO architecture, to enable speech interaction, we have an avatar that uses the speaker worker (discussed in details below) to talk to the user when needed. Depending on the environment configuration, the transcription microphones might hear itself talking and transcribe its own utterances. To avoid this problem, the transcript worker pauses audio capturing on the far-range microphones, but keep the close-range microphones open so that users can still talk to the room avatar when it is talking.

An intelligent space may have multiple transcription workers running at the same time, and thanks to the centralized message broker, this is easily supported by the CELIO framework. In fact, we provide a webpage portal that allows users to transcribe the audio from their own devices such as laptops and mobile phones. This greatly expands users' access to the intelligent space.

*Vision Workers*
The CELIO architecture includes vision workers that provide computer vision sensing of the intelligent space. So far, we have created vision workers that export the Kinect[6] data and publishes them to the message broker. The data published by the Kinect worker includes the 3D coordinates of the users' body joints, the state of the hands (opened or closed), the orientation of the users' heads, and some facial features such as smile and eyes closed or opened. The main function of the Kinect data is to enable users to point to contents and objects in the space and interact with them just with their arms. From the Kinect body joint data, we take the vector from the elbow to the hand as the pointing vector, and publishes the pointing vectors to the message broker. This way, display workers or other agents can subscribe to these pointing events and respond to them as needed. Another use of the Kinect data is to simply detect the location of the users, and to tailor the display and speak output to the users based on their locations.

---

[5] http://www.ibm.com/watson/developercloud/speech-to-text.html

[6] https://developer.microsoft.com/en-us/windows/kinect

We have also created a face recognition REST service based on the OpenFace library[7]. The Kinect vision worker is connected to the face recognition service such that when a user with an unknown identity is detected and he or she is facing the Kinect camera, the worker captures the face image and send it to the face recognition service. When an identity is returned, it is incorporated into the data stream that the Kinect worker publishes. This again helps multimodal fusion and allow us to associate a user's speech and his or her gestures via the user identity.

*Pointer Workers*

The pointer workers provide pointing data for the entire space. Currently, we have created several workers that support absolute pointing devices. The pointing location of an absolute pointing device is the intersection between a receiving surface and the ray casted in the direction of the pointer. For example, a laser pointer is an absolute pointing device, though its pointing location needs to be analyzed from camera images [12]. Mice, by contrast, is a relative pointing device because it controls a cursor's movement, not its locations. In a large space with large displays in many directions, it is found that absolute pointing may be more effective than relative pointing because relative pointers often require too many clutching actions (moving the mouse back to a nearby location to start a new cursor movement) [18]. In addition, with absolute pointing, we can designate non-display surfaces that also respond to pointing events for controlling objects. For example, we can designate a lightbulb on the ceiling as an interactive surface, and the user can then point to the lightbulb and click a button on the pointer to turn it on and off. With relative pointing devices, it may not be clear to the user how the cursor controlled by the device would move across the space to places like the ceiling, especially when there is no visual feedback when the cursor moves outside a display.

We have created workers to support multiple kinds of absolute pointing devices. One device we use in our lab is a "wand" from the Oblong Industry. It requires a grid of ultrasound emitters installed on the ceiling to provide location sensing for the wand. The wand uses its onboard ultrasound microphones to trilaterate its location and orientation in space, thereby obtaining its precise pointing location on the displays. We have also adapted a consumer-grade Virtual Reality technology as our absolute pointing device. The VR product we have adapted is called HTC Vive. It uses infrared laser scanning stations to provide reference signals to the controllers. The controllers have infrared sensors, and by calculating the time difference between different reference signals, it calculates its orientation and location in space. Finally, as previous discussed, we also extract pointing data from the Kinect data, using users' arms as pointers.

To support these different types of pointing devices, we defined a common, but extensible pointer data format. The pointer data are encoded in JSON format and have the following required fields:

- **loc**. A three-element array containing the pointer's coordinates in millimeters.
- **aim**. A three-element array containing the pointer's direction as a unit vector.
- **buttons**. An array containing the labels of the buttons that are pressed.
- **type**. A string identifies the type of the pointer such as Kinect, wand, or vive.
- **name**. A unique identifier for the pointer.
- **details**. An object containing other data that belong to this particular type of pointers.

The first three fields provide the essential information for the event subscribers to determine where the pointing location is and what buttons were pressed. Other data pertaining to a specific type of pointer is also in the "details" field and can be used by the applications if they choose to.

**ACTUATORS**

Actuator workers can be thought of the things that react to events and commands in the space. They are hardware units that are typically internet of things and can be accessed and controlled over the network, for instance, REST. To support spatial interaction with these actuators, we use a generic hotspot to interpret the pointing events. In addition, these actuators also act based on received commands and emit change events to the message broker.

**Hotspot**

Hotspots are regions in the space that will receive pointing events from different input workers such as 3D pointing, gestures, and touch. These hotspots use the interactive space's spatial reference framework to define its 3D-bounding box. Hotspots are defined in the workers that manages their corresponding things, displays and actuators such as speaker, lighting and camera. On pointing, a hotspot generates events such as enter, leave, button_<state>, gesture_<type>, move, dragStart, dragEnter, dragLeave, and dragEnd. For instance, a user can pick twitter sentiment data on the visualization and drop on the RGB LED lamp. The Lamp worker receives the dragDrop event with sentiment data and changes its color to red or green. Figure 2 illustrates the hotspots around things and displays in the interactive space and shows drag and drop of events from the display wall A to the lamp.

---

[7] https://cmusatyalab.github.io/openface/

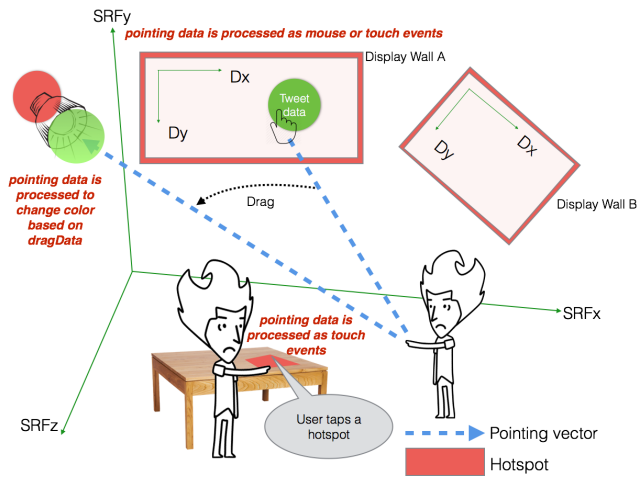

**Figure 2. Hotspots in the interactive space defined for display walls and things. User drags a twitter data and drop on the RGB LED lamp and lights up as green to reflect the sentiment of the data.**

All hotspots hit by a pointing vector receive point of intersection, distance along the hotspot normal and distance along the pointing vector. Once a hotspot receives the pointing event, it can translate into its own reference coordinate system and pass the pointing events to its children or directly emit events such as tap, swipe, grab, click, dragStart and dragEnd. Our system emits both native events as well as special event. At display worker, hotspot pointing events are first emitted as mouse and touch events along with additional event details. For instance, a Wand has 3 buttons and Vive controller has a grip button, a menu button, a system button, and a trackpad. In addition to translating the events from these devices as a typical mouse and touch events, these details are embedded in the event data. So, the application can be written both for Web and Interactive Spaces without any additional development overhead. In case, the developers want to enrich the event processing, they can process the raw data embedded in the pointing event streams. This event flow is illustrated in Figure 3.

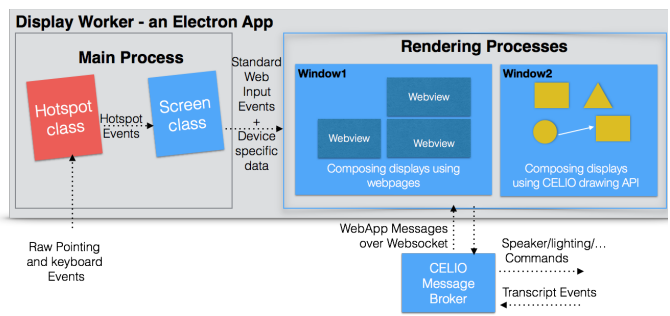

---

[8] http://electron.atom.io

**Figure 3. Display Worker – an Electron™ app.** The event flow from pointing devices are captured by the hotspot class. The screen class transforms the pointing data from 3d to 2d display wall coordinate system and sends as standard web input events into webpages and web objects drawn using CELIO drawing package. In addition, the web pages can benefit from interacting with the message broker to which the display worker is connected to in the space.

**Display Worker**

Display worker manages the contents on displays in the space. Displays are usually built as continuous display walls with bezel, curved immersive walls, distributed in space or even mounted on a robot arm for on-demand re-orientation. One of the design goal is to enable Web UI developers to quickly use our framework to develop applications for the interactive spaces. We use Electron[8] – a cross-platform desktop application development framework based on web technologies such as HTML, JavaScript and CSS to build the display worker. The advantages of using Electron framework include loading large media files from local system on webpages, piping custom events into webpages, programmatically move and resize browser windows, cross-browser window communication using IPC, use web animation API for even moving browser window and its contents, rendering process separation from main process and so on. We adapted the Electron framework to enable creation of content in the following ways:

1. Users can create and load multiple Electron BrowserWindows and load required content in them

2. Users can open a template file (an HTML file with containers laid out using CSS methods such as FlexBox and floats) within a BrowserWindow. They can on-demand insert Electron WebView elements into the template file. Webpages loaded within Electron WebView runs in a separate process and optionally can communicate with the parent window. With this support, they can now apply Web Animation

API[9] to create great visual effects for enhancing content visualization experience

3. Users can create a canvas and draw shapes and text using CELIO drawing API. At the moment, we support basic HTML5 Canvas drawing. In future, we plan to support d3 and three.js

We organize the web element (div) layers stacked one below the other within a BrowserWindow using a fixed template file as follows:

1. Cursor (Top) layer – this layer is used to draw multi-user cursor. Each pointing device and its events will be represented using custom cursor images. Mouse cursor are drawn by Operating System and are not replicated in this layer.

2. Canvas layer – this layer is used for rendering user sketches – shapes, highlights and notes. Shaping sketching and notes placement are done using pointing device. This layer can be programmatic rendered using CELIO drawing API

3. Content layer – this layer is used to load user template files.

4. Background (Bottom) layer – this layer is used for ambient visualization

**THE CELIO APPLICATION PROGRAM INTERFACE**

As described previously, in the CELIO framework, the applications communicate with the workers by subscribing to events, sending commands to the message broker, and making direct REST calls. Since the applications subscribe to events by setting a topic name, one of the main jobs we need to do in defining the CELIO API is to define the topic name convention. We follow a general convention used in RabbitMQ that a topic name contains multiple words separated by a dot such as "close-range.final.transcript". On top of this convention, we require that the topics published by the workers contain qualifiers from more specific type designation to more general type designation. For example, in "close-range.final.transcript", the last word, "transcript", is the most general type designation. The second word specifies that this message is a final result of the transcription of one sentence instead of an interim result. The first word specifies that the result comes from a close-range microphone. With this naming convention, applications can easily filter out events using wildcard matching. For example, "*.final.transcript" will only subscribe to all final transcripts, whereas "*.*.transcript" subscribes to all transcripts including interim transcripts.

Though applications can subscribe to the events and publish events directly use RabbitMQ libraries available in many languages, we are providing a higher level API so that the developers do not need to know about RabbitMQ or the topic names with which the workers publish their events. Currently, we provide such an API for Javascript. Below is a snippet of the use of the API for subscribing to transcripts and making the speaker say a sentence:

```
var io = new CELIO();
var transcript = io.getTranscript();
transcript.onFinal(function(msg)
{//process the message});
var speaker = io.getSpeaker();
speaker.speak("Hello",     {location:
"Front", voice: "Lisa"});
```

As can be seen, the application developers do not need to know anything about how to connect to RabbitMQ and which topic to subscribe to. These details are all encapsulated within those function calls.

**USE CASES**

We used the CELIO architecture to create several applications that take advantage of the multimodal input available in the architecture. In one application, we turned the pointer devices into universal remotes by creating hotspots for actuators that are connected to the network. The users can then use a pointer to turn on/off a recording sign, change speaker volume with a sliding gesture, and control pan-tilt-zoom cameras. In particular, in the camera control application, the user uses speech commands to ask the camera to look at the user, look at where the user is pointing at, or take a picture and post it on the screen. This concept of universal remote is reminiscent of the previous idea of magic wand [20], but with CELIO's architectural support, such applications are very easy to develop and they automatically support different pointing devices because a hotspot object subscribes to all absolute pointing events.

In another application, we created a US map visualization that responds to the user's movement and distance from the screen. When the user is far away from the screen, only state-level data are shown to reduce clutter. As the user moves closer to the screen, county-level data are revealed. The user can also point to a state and say "zoom in" to enlarge a state, or say "zoom out" to go back to see the entire US map. Again, the CELIO architecture make user location data and speech data easily accessible, which facilitated the creation of such multimodal applications.

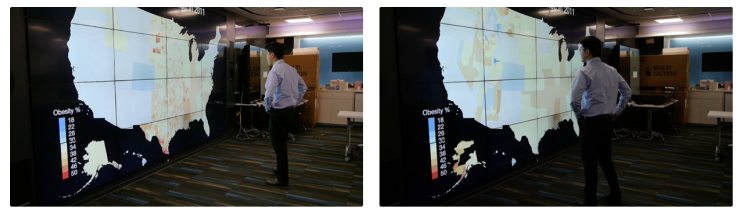

(a)　　　　　　　　　　　(b)

---

[9] https://w3c.github.io/web-animations/

Figure 4 (a) and (b) shows county level data based on a user's proximity to the display. The proximity data is consumed directly by the webapp from the message broker and updates the visualization based on user movement.

**CONCLUSION AND FUTURE WORK**

In this paper, we presented the CELIO (Cognitive Environment Library for Input/Output) application development framework for interactive spaces. The framework is designed based on system design requirements that were derived from lesson learned from real world interactive spaces application. These requirements highlighted the need for a distributed micro service architecture with support for spatial awareness of things, displays, devices and users; multi-user interactions; inter-space communication; and interaction, in addition to, the usual low latency and plug and play capabilities. The CELIO application development framework caters to these requirements using a spaces registry federation, distributed message broker, input workers, actuators, and display worker. Finally, we demonstrated the use of the CELIO Application Program Interface by illustrating development of universal remote and proxemics experiences in our lab. In future, we need to address distributed visualization capabilities across displays separated in space, global state space and conditional action dispatching (planning) capabilities to support intelligent rooms.